\documentclass[aps,preprint]{revtex4}%
\usepackage{amsfonts}
\usepackage{amsmath}
\usepackage{amssymb}
\usepackage{graphicx}%
\providecommand{\U}[1]{\protect\rule{.1in}{.1in}}
\newtheorem{theorem}{Theorem}
\newtheorem{acknowledgement}[theorem]{Acknowledgement}

\begin{document}
\preprint{ }
\title{Massive Gravity: Resolving the Puzzles\\ }
\author{Lasma Alberte}
\affiliation{Theoretical Physics, Ludwig Maxmillians University,Theresienstr. 37, 80333
Munich, Germany}
\author{Ali H. Chamseddine}
\affiliation{American University of Beirut, Physics Department, Beirut, Lebanon, and
I.H.E.S. F-91440 Bures-sur-Yvette, France}
\author{Viatcheslav Mukhanov}
\affiliation{Theoretical Physics, Ludwig Maxmillians University,Theresienstr. 37, 80333
Munich, Germany and Department of Physics, New York University, NY 10003, USA}
\pacs{PACS number}

\begin{abstract}
We consider the massless limit of Higgs gravity, where the graviton becomes
massive when the scalar fields acquire expectation values. We determine the
Vainshtein scale and prove that massive gravity smoothly goes to General
Relativity below this scale. We find that the Vainshtein scale depends on the
particular action of scalar fields used to give mass to the graviton.

\end{abstract}
\maketitle

\section{Introduction}

In the recent paper \cite{mukh} we (A.Ch.,V.M) have proposed a Higgs mechanism
for gravity. In our model the graviton becomes massive as a result of
spontaneous symmetry breaking, where four scalar fields acquire non-vanishing
expectation values. As a result, three out of four degrees of freedom of
scalar fields are absorbed producing a massive graviton with five degrees of
freedom, while one degree of freedom remains strongly coupled. Our model is
explicitly diffeomorphism invariant and, in distinction from bigravity
theories, it is simply given by General Relativity supplemented with the
action of four extra scalar fields. Therefore it is completely analogous to
the standard Higgs mechanism used to give masses to the gauge fields, where
masses are acquired as a result of the interaction with external classical
scalar fields. For instance, in the standard electroweak theory one also uses
four (real) scalar fields to give masses to three vector bosons, and one
remaining degree of freedom becomes a Higgs boson. However, in distinction
from electroweak theory, in our case the analogue of the Higgs boson remains
strongly coupled and hence completely decouples from gravity and other matter.

The theory with four scalar fields was exploited before by several authors
(see \cite{thooft,kaku,dub} and references therein). In our case we have found
the Lagrangian which resolved the problems that faced finding a consistent
theory for massive gravitons. On one hand the model produces a graviton mass
term with explicitly invariant form even for finite diffeomorphisms, and on
the other hand, keeps the dangerous mode which could produce a ghost, in the
strong coupling regime where it is completely harmless. In the linear order
the mass term is of the Fierz-Pauli form \cite{pauli}, which is uniquely fixed
by the requirements of the absence of extra scalar degree of freedom. The
analysis by Deser and Boulware \cite{boul} however lead to the conclusion that
in the massive theory the extra scalar degree of freedom reappears at
nonlinear level and does not decouple, thus making massive gravity to be an
ill-behaved theory. In distinction from \cite{boul}, where diffeomorphism
invariance is explicitly spoiled, our theory is diffeomorphism invariant and
therefore the $g_{0\alpha}$ components of the metric remain always the
Lagrange multipliers, while as we will show later, the scalar fields are
always in the strong coupling regime above so called Vainshtein energy scale.
This corresponds to extremely small energy and therefore the possible ghost is
irrelevant. 

There were many interesting attempts to extend massive gravity beyond the
linear approximation in a way where one can avoid the extra mode and ghost,
also at the nonlinear level (see, for instance,
\cite{dgp,gabad,rahm,gruzinov,gr1,gr} and references there). In particular, in
the recent interesting papers \cite{gr1,gr} it was found an extension of the
Fierz-Pauli action for which the ghosts are absent even at nonlinear level at
the decoupling limit. 

The main purpose of this paper is to investigate the existence of a smooth
limit of our model to Einstein gravity, when the mass of the graviton
vanishes. It was noticed long ago by van Dam, Veltman and Zakharov
\cite{dam,zak} that in linearized massive gravity the extra scalar mode of the
graviton did not disappear and remained coupled to matter even in the limit of
a vanishing graviton mass. In turn, this spoils predictions of General
Relativity either for the perihelion precession or deflection of starlight.
This effect is known as the van Dam-Veltman-Zakharov (vDVZ) discontinuity and
was first thought to be a no-go theorem for massive theories of gravity
\cite{dam,zak}. However, it was pointed out by Vainshtein that the
discontinuity could be an artifact due to the breakdown of the perturbation
theory of massive gravity in the massless limit \cite{vain}. He has shown that
in the case of gravitational field produced by a source of mass $M_{0}$ the
nonlinear corrections become important at scales $r<R_{V}\equiv M_{0}%
^{1/5}m_{g}^{-4/5}$ (in Planck units) and conjectured that in the strong
coupling regime General Relativity is restored. When the mass of the graviton
$m_{g}$ vanishes the Vainshtein radius $R_{V}$ grows and becomes infinite,
thus providing a continuous limit to General Relativity in case the Vainshtein
conjecture is correct. At distances $r\ll R_{V},$ around a static spherically
symmetric massive source of mass $M_{0}$ the full non-linear strongly coupled
massive gravity has to be considered in order to recover the Einstein theory,
which makes the proof of the Vainshtein conjecture non trivial. The question
of continuous matching of the solutions below and above the Vainshtein radius
have been extensively addressed in recent literature. The first model where
such a transition was demonstrated is Dvali-Gabadadze-Porrati (DGP) model
which imitates many features of massive gravity\cite{dgp,dvali}. There was a
claim that in the bigravity version of massive graviton the corresponding
solutions do not match \cite{damour}, but it was recently shown that this
claim is not justified \cite{bab, bab2, bab3}.

In this paper we will find the Vainshtein scale and will prove Vainshtein
conjecture in the Higgs model of massive gravity in the case when the
gravitational field is produce by a source of mass $M_{0}.$ Moreover, we will
find how the concrete value of the Vainshtein scale depends on the nonlinear
extension of the Pauli Fierz term, or in other words on the interactions of
scalar fields used to produce massive gravity. As a result we will determine
possible Vainshtein scales for a wide class of Higgs gravity models. We will
also derive in our model the leading corrections to the gravitational
potential within Vainshtein scale, which are similar, but not identical to
this type of correction obtained in the framework of the DGP model in
\cite{dvali,gruzinov,dvali2}.

Finally, we will discuss the implications of our results obtained in classical
theory when extended to quantum theory. In particular we argue that in quantum
theory there must be a cutoff scale at energies $m_{g}^{4/5},$ above which the
scalar fields enter strong coupling regime and completely decouple from
gravity and other matter. Because this scale is extremely small for the
realistic mass of the graviton it makes the problem of ghost which could
appear only below this scale completely irrelevant. For the scalar and vector
modes of the massive graviton the cutoff scale is an analog of the Planckian
scale for the tensor graviton modes, which also become strongly coupled above
Planck scale. The obtained cutoff scale is in agreement with results of
\cite{dvali,arkani,crim}.

\section{Higgs for graviton: Basics}

We employ four scalar fields $\phi^{A},A=0,1,2,3$ to play the role of Higgs
fields. These will acquire a vacuum expectation value proportional to the
space-time coordinates, thus giving mass to the graviton. Let us introduce the
\textquotedblleft composite metric\textquotedblright%
\begin{equation}
H^{AB}=g^{\mu\nu}\partial_{\mu}\phi^{A}\partial_{\nu}\phi^{B}, \label{1}%
\end{equation}
which is scalar with respect to diffeomorphism transformations. The field
indices $A,B,\cdots,$ are raised and lowered with the Minkowski metric
$\eta_{AB}$. The diffeomorphism invariant action which will be used as our
model, is given by
\begin{equation}
S=-\frac{1}{2}\int d^{4}x\,\sqrt{-g}R+\frac{M^{2}}{8}\int d^{4}x\,\sqrt
{-g}\left[  3\left(  \left(  \frac{1}{4}H\right)  ^{2}-v^{2}\right)
^{2}-v^{2}\tilde{H}_{B}^{A}\tilde{H}_{A}^{B}\right]  , \label{2}%
\end{equation}
where
\begin{equation}
\tilde{H}_{B}^{A}=H_{B}^{A}-\frac{1}{4}\delta_{B}^{A}H, \label{3}%
\end{equation}
is the traceless part of the \textquotedblleft composite
metric\textquotedblright\ and where we have set $8\pi G=1.$ The parameter $v$
controls the symmetry breaking scale. As will be seen later, the induced mass
of the graviton is equal to $m_{g}=Mv^{2}$ and hence when $v\rightarrow0$
gravity becomes massless. It is clear that in this limit the only surviving
term in action (\ref{2}) is Einstein gravity and $M^{2}H^{4}$ for the four
scalar fields, which are in the regime of strong coupling and do not possess
linear propagators. In the phase with restored symmetry the total number of
degrees of freedom is six: two of them describe massless graviton and four
correspond to scalar fields which are decoupled from gravity at linear level.

We show next that when the symmetry is broken, three out of four scalar fields
are \textquotedblleft eaten\textquotedblright\ and produce the massive
graviton with five degrees of freedom, while the \textquotedblleft
surviving\textquotedblright\ degree of freedom will remain strongly coupled.
In case when $v\neq0,$ the unique Minkowski vacuum solution of the equations
of motion, $g_{\mu\nu}=\eta_{\mu\nu},$ corresponds to the fields, which
linearly grow with coordinates, that is, $\phi^{A}=\sqrt{v}\delta_{\beta}%
^{A}x^{\beta}$. Let \ us consider perturbations around Minkowski background,
\begin{equation}
g^{\mu\nu}=\eta^{\mu\nu}+h^{\mu\nu},\text{ \ \ \ \ \ }\phi^{A}=\sqrt{v}\left(
x^{A}+\chi^{A}\right)  \label{4}%
\end{equation}
and define
\begin{align}
\bar{h}_{B}^{A}  &  \equiv\frac{1}{v}H_{B}^{A}-\delta_{B}^{A}=h_{B}%
^{A}+\partial^{A}\chi_{B}+\partial_{B}\chi^{A}\label{5}\\
&  +\partial_{C}\chi^{A}\partial^{C}\chi_{B}+h_{C}^{A}\partial^{C}\chi
_{B}+h_{B}^{C}\partial_{C}\chi^{A}+h_{D}^{C}\partial^{D}\chi_{B}\partial
_{C}\chi^{A},\nonumber
\end{align}
where indices are moved with the Minkowski metric, in particular, $\chi
_{B}=\eta_{BC}\chi^{C}$ and $h_{B}^{A}=\eta_{BC}\delta_{\mu}^{A}\delta_{\nu
}^{C}h^{\mu\nu}.$ We point out that we have included a factor $\sqrt{v}$ as
coefficient of $\chi^{A}$ to obtain simpler expressions. In reality in all our
results that will subsequently follow we have to make the replacement%
\[
\chi^{A}\rightarrow\chi^{A}\frac{1}{\sqrt{v}}=\left(  \frac{M}{m_{g}}\right)
^{\frac{1}{4}}\chi^{A}.
\]
This, however, will not effect most of our conclusions, and we will thus
comment on it only when necessary. With the help of the expressions
\[
H=v\left(  \bar{h}+4\right)  ,\text{ \ \ }\tilde{H}_{B}^{A}\tilde{H}_{A}%
^{B}=v^{2}\left(  \bar{h}_{B}^{A}\bar{h}_{A}^{B}-\frac{1}{4}\bar{h}%
^{2}\right)  ,
\]
we can rewrite the action for the scalar fields in the following form
\begin{equation}
S_{\phi}=\frac{M^{2}v^{4}}{8}\int d^{4}x\,\sqrt{-g}\left[  \bar{h}^{2}-\bar
{h}_{B}^{A}\bar{h}_{A}^{B}+\frac{3}{4^{2}}\bar{h}^{3}+\frac{3}{4^{4}}\bar
{h}^{4}\right]  . \label{6}%
\end{equation}
We would like to stress that we did not use any approximations to derive
(\ref{6}), and $\bar{h}_{B}^{A}$ are diffeomorphism invariant combinations of
the scalar fields and metric up to an arbitrary order.

\section{Physical degrees of freedom of the massive graviton}

We consider now small perturbations of the metric and scalar fields and
neglect higher order terms. In this case
\begin{equation}
\bar{h}_{B}^{A}=h_{B}^{A}+\partial^{A}\chi_{B}+\partial_{B}\chi^{A}%
+\mathcal{O}(h^{2},\chi^{2}), \label{7}%
\end{equation}
and in the leading order, action (\ref{6}) describes Fierz-Pauli massive
gravity, where the mass of the graviton is equal to $m_{g}=Mv^{2}.$ However,
we have to stress that in distinction from the Fierz-Pauli theory our model
does not break diffeomorphism invariance and coincides with this theory only
in the unitary gauge where all $\chi^{A}=0.$ In turn, imposing these gauge
conditions completely fixes the coordinate system making the interpretation of
the results rather obscure. If one would try to treat $\chi^{A}$ as
St\"{u}ckelberg \textquotedblleft vector\textquotedblright\ field and consider
\textit{the diffeomorphism transformations for the vectors} \ rather than some
obscure \textquotedblleft fictitious\textquotedblright\ symmetries, then one
unavoidably would conclude that the \textquotedblleft vector
components\textquotedblright\ \textit{must} be treated as the perturbations of
four scalar fields with nonzero background values, thus arriving at our model.
As we will see in the next section the difference between the noncovariant
Fierz-Pauli approach and our model becomes even more dramatic at higher
orders. However, we first study the linearized theory using Lorentz-violating
approach to explicitly reveal the true physical degrees of freedom of the
massive graviton. Namely, we use the method usually applied in cosmological
perturbation theory and classify the metric perturbations according to the
irreducible representations of the spatial rotation group \cite{vm}. The
$h_{00}$ component of the metric behaves as a scalar under these rotations and
hence%
\begin{equation}
h_{00}=2\phi, \label{8}%
\end{equation}
where $\phi$ is a 3-scalar. The space-time components $h_{0i}$ can be
decomposed into a sum of the spatial gradient of some 3-scalar $B$ \ and a
vector $S_{i}$ with zero divergence:%
\begin{equation}
h_{0i}=B_{,i}+S_{i}, \label{9}%
\end{equation}
where $B_{,i}=\partial B/\partial x^{i}=\partial_{i}B$ and $\partial^{i}%
S_{i}=0.$

In a similar way $h_{ij}$ can be written as
\begin{equation}
h_{ij}=2\psi\delta_{ij}+2E_{,ij}+F_{i,j}+F_{j,i}+\tilde{h}_{ij}, \label{10}%
\end{equation}
where $\partial^{i}F_{i}=0$ and $\partial^{i}\tilde{h}_{ij}=0=\tilde{h}%
_{i}^{i}.$ The irreducible \textit{tensor perturbations} $\tilde{h}_{ij}$ have
two independent components and describe the graviton with two degrees of
freedom in a diffeomorphism invariant way. The \textit{scalar perturbations}
are characterized by the four scalar functions $\phi,\psi,B,$ and $E.$ In
empty space they vanish and are induced entirely by matter, which in our case
are\ the scalar fields. The \textit{vector perturbations} of the metric
$S_{i}$ and $F_{i}$ are also due to the matter inhomogeneities The matter
perturbations can also be decomposed into scalar and vector parts:%
\begin{equation}
\chi^{0}=\chi^{0},\text{ }\chi^{i}=\tilde{\chi}^{i}+\pi_{,i} \label{11}%
\end{equation}
where $\partial_{i}\tilde{\chi}^{i}=0$. In the linear approximation, scalar,
vector and tensor perturbations are decoupled and can be analyzed separately.

\textbf{Scalar perturbations. }Up to first order in perturbations we have
$h^{\alpha\beta}=-\eta^{\alpha\nu}\eta^{\beta\mu}h_{\mu\nu}$ and using the
definition of $\bar{h}_{B}^{A}$ in (\ref{5}) we find that in the leading order
approximation%
\begin{equation}
^{(S)}\bar{h}_{0}^{0}=-2\phi+2\dot{\chi}^{0},\text{ }^{(S)}\bar{h}_{i}%
^{0}=-B_{,i}-\dot{\pi}_{,i}+\chi_{,i}^{0},\text{\ \ }^{(S)}\bar{h}_{k}%
^{i}=2\psi\delta_{ik}+2E_{,ik}+2\pi_{,ik}. \label{12}%
\end{equation}
Substituting these expressions in (\ref{6}), keeping only second order terms,
and expanding the Einstein action up to second order in metric perturbations
we obtain the following action for the scalar perturbations:%
\begin{align}
^{(S)}\delta_{2}S  &  =\int d^{4}x\,\left\{  -3\dot{\psi}^{2}+\psi_{,i}%
\psi_{,i}+\phi\left[  2\Delta\psi-m_{g}^{2}(3\psi+\Delta(E+\pi))\right]
\right. \nonumber\\
&  +2\dot{\psi}\Delta\left(  B-\dot{E}\right)  +m_{g}^{2}\left[  3\psi\left(
\psi+\dot{\chi}^{0}\right)  +\left(  2\psi+\dot{\chi}^{0}\right)  \Delta
(E+\pi)\right. \nonumber\\
&  \left.  \left.  +\tfrac{1}{4}\left(  \chi^{0}-B-\dot{\pi}\right)
_{,i}\left(  \chi^{0}-B-\dot{\pi}\right)  _{,i}\right]  \right\}  , \label{13}%
\end{align}
where $m_{g}^{2}=M^{2}v^{4}$ and the dot denotes derivative with respect to
time. We see that $\phi$ is a Lagrangian multiplier which implies the
constraint%
\begin{equation}
\Delta\psi=\frac{m_{g}^{2}}{2}(3\psi+\Delta(E+\pi)). \label{14}%
\end{equation}
Another constraint is obtained by variation with respect to $B$:%
\begin{equation}
\dot{\psi}=-\frac{m_{g}^{2}}{4}\left(  \chi^{0}-B-\dot{\pi}\right)  .
\label{15}%
\end{equation}
To simplify further the calculations we select the longitudinal gauge $B=E=0,$
which when used in conjunction with (\ref{14}), simplifies the action
(\ref{13}) to%
\begin{align*}
^{(S)}\delta_{2}S  &  =\int d^{4}x\,\left[  -3\dot{\psi}^{2}+\psi_{,i}%
\psi_{,i}\right. \\
&  \left.  +m_{g}^{2}\left(  3\psi\left(  \psi+\dot{\chi}^{0}\right)  +\left(
2\psi+\dot{\chi}^{0}\right)  \Delta\pi+\tfrac{1}{4}\left(  \chi^{0}-\dot{\pi
}\right)  _{,i}\left(  \chi^{0}-\dot{\pi}\right)  _{,i}\right)  \right]  .
\end{align*}
Using constraints (\ref{14}) and (\ref{15}) with $B=E=0,$ imply%
\begin{align}
m_{g}^{2}\Delta\pi &  =\left(  2\Delta-3m_{g}^{2}\right)  \psi\label{17a}\\
m_{g}^{2}\Delta\chi^{0}  &  =-\left(  2\Delta+3m_{g}^{2}\right)  \dot{\psi}
\label{17b}%
\end{align}
which can be inverted to express $\pi$ and $\chi^{0}$ in terms of $\psi$:%
\begin{equation}
\pi=\left(  \frac{2}{m_{g}^{2}}-\frac{3}{\Delta}\right)  \psi, \label{18a}%
\end{equation}%
\begin{equation}
\chi^{0}=-\left(  \frac{2}{m_{g}^{2}}+\frac{3}{\Delta}\right)  \dot{\psi}
\label{18b}%
\end{equation}
Substituting these relations in the action above we obtain%
\begin{align}
^{(S)}\delta_{2}S  &  =\int d^{4}x\,\left[  -3\dot{\psi}^{2}+\psi_{,i}%
\psi_{,i}+m_{g}^{2}\left(  \frac{6}{m_{g}^{2}}\dot{\psi}^{2}-\frac{4}%
{m_{g}^{2}}\psi_{,i}\psi_{,i}-3\psi^{2}\right)  \right] \nonumber\\
&  =-3\int d^{4}x\,\left[  \psi\left(  \partial_{t}^{2}-\Delta+m_{g}%
^{2}\right)  \psi\right]  . \label{19}%
\end{align}
Note that the potential $\psi$ is gauge invariant with respect to
infinitesimal diffeomorphism transformations: $x^{\alpha}\rightarrow\tilde
{x}^{\alpha}=x^{\alpha}+\xi^{\alpha}$. Therefore the derived result does not
depend on the particular gauge we used to simplify the calculations of the
action. First of all we see that the scalar mode which was non-propagating in
the absence of the scalar fields has become dynamical. The variable
$u=\sqrt{6}\psi$ is the canonical quantization variable for the scalar degree
of freedom of metric perturbations. It is entirely induced by perturbation of
the scalar fields $\pi$ and $\chi^{0}$. In the linear approximation we have to
be careful in taking the limit $m_{g}\rightarrow0$ because of the inverse mass
dependence in the relations (\ref{18a}) and (\ref{18b}). In reality we have to
consider instead equations (\ref{17a}) and (\ref{17b}) which implies that
$\psi=0$ as in the vacuum case. Thus the famous vDVZ discontinuity
\cite{dam,zak} \ is not present. In addition, as mentioned before, when taking
the limit $m_{g}\rightarrow0$ we have to replace the fields $\pi$ and
$\chi^{0}$ with $\left(  \frac{M}{m_{g}}\right)  ^{\frac{1}{4}}\pi$ and
$\left(  \frac{M}{m_{g}}\right)  ^{\frac{1}{4}}\chi^{0}$ but this leads to the
same result that $\psi=0.$ We note, however, that in the $m_{g}\rightarrow0$
the Higgs action reduces to the $M^{2}H^{4}$ term, and there are higher order
non-linear contributions to $\psi.$ In the next section we will show that
above a certain energy scale the scalar mode ceases to propagate and becomes
confined due to nonlinear corrections to the equations. As a result the vDVZ
discontinuity is avoided completely and we obtain a smooth limit to General
Relativity when symmetry is restored and the graviton becomes massless.

\textbf{Vector perturbations.} For the vector perturbations
\begin{equation}
^{(V)}\bar{h}_{i}^{0}=-S_{i}-\dot{\tilde{\chi}}^{i},\text{ \ }^{(V)}\bar
{h}_{k}^{i}=F_{i,k}+F_{k,i}+\tilde{\chi}_{,i}^{k}+\tilde{\chi}_{,k}^{i}.
\label{20}%
\end{equation}
Up to second order in perturbations the action for the vector modes is
\begin{align}
^{(V)}\delta_{2}S  &  =\frac{1}{4}\int d^{4}x\,\left[  \left(  \dot{F}%
_{i}-S_{i}\right)  _{,k}\left(  \dot{F}_{i}-S_{i}\right)  _{,k}\right.
\nonumber\\
&  \left.  +m_{g}^{2}\left(  \left(  \dot{\tilde{\chi}}^{i}+S_{i}\right)
\left(  \dot{\tilde{\chi}}^{i}+S_{i}\right)  -\left(  F_{i}+\tilde{\chi}%
^{i}\right)  _{,k}\left(  F_{i}+\tilde{\chi}^{i}\right)  _{,k}\right)
\right]  . \label{21}%
\end{align}
Variation of this action with respect to $S_{i}$ gives the constraint equation%
\[
\Delta\left(  \dot{F}_{i}-S_{i}\right)  =-m_{g}^{2}\left(  \dot{\tilde{\chi}%
}^{i}+S_{i}\right)  ,
\]
which allows us to express $S_{i}$ as%
\begin{equation}
S_{i}=\frac{1}{\Delta-m_{g}^{2}}\left(  \Delta\dot{F}_{i}+m_{g}^{2}%
\,\dot{\tilde{\chi}}^{i}\right)  . \label{22}%
\end{equation}
Substituting this expression into (\ref{21}) we obtain
\begin{equation}
^{(V)}\delta_{2}S=-\frac{1}{2}\int d^{4}x\,\frac{m_{g}^{2}\Delta}{2\left(
\Delta-m_{g}^{2}\right)  }\left[  \left(  F_{i}+\tilde{\chi}^{i}\right)
\left(  \partial_{t}^{2}-\Delta+m_{g}^{2}\right)  \left(  F_{i}+\tilde{\chi
}^{i}\right)  \right]  . \label{23}%
\end{equation}
In the limit $m_{g}\rightarrow0$ the action for the vector modes vanishes even
after replacing $\tilde{\chi}^{i}\rightarrow\left(  \frac{M}{m_{g}}\right)
^{\frac{1}{4}}\tilde{\chi}^{i}.$ The canonical gauge invariant quantization
variable in this case is the 3-vector
\begin{equation}
V^{i}=\sqrt{\frac{m_{g}^{2}\Delta}{2(\Delta-m_{g}^{2})}}\left(  F_{i}%
+\tilde{\chi}^{i}\right)  , \label{24}%
\end{equation}
which describes two physical degrees of freedom as this vector satisfies an
extra condition $\partial_{i}V^{i}=0.$

\textbf{Tensor perturbations. }For the tensor perturbations the result is
straightforward%
\begin{equation}
^{(T)}\delta_{2}S=-\frac{1}{8}\int d^{4}x\left[  \tilde{h}_{ij}\left(
\partial_{t}^{2}-\Delta+m_{g}^{2}\right)  \tilde{h}_{ij}\right]  . \label{25}%
\end{equation}
This action describes the pure gravitational degrees of freedom which have
become massive. Because $\tilde{h}_{ij}$ satisfies four extra conditions
$\partial^{i}\tilde{h}_{ij}=0=\tilde{h}_{i}^{i}$ \ the tensor perturbations
have two physical degrees of freedom.

Thus, we have decomposed the massive graviton with five degrees of freedom
into physical gauge invariant components: a scalar part $\psi$ (with one
degree of freedom), a vector part $V^{i}$ (2 degrees of freedom) and a tensor
part $\tilde{h}_{ij}$ (2 degrees of freedom). After quantization they acquire
their independent gauge invariant propagators.

The metric components are the subject of minimal vacuum quantum fluctuations.
In particular, the amplitude of the vacuum fluctuations of $\psi$ and
$\tilde{h}_{ij}$ at scales $\lambda\ll1/m_{g}$ $\ $ are about%
\[
\psi\sim\tilde{h}_{ij}\sim\frac{1}{\lambda},
\]
in Planck units. They become of the order of one at the Planck scale
$l_{\mathrm{Pl}}\simeq10^{-33}$ $\mathrm{cm}$ where non perturbative quantum
gravity becomes important. The amplitude of the vector vacuum metric
fluctuations is much smaller. In fact, for $\lambda\ll1/m_{g},$ their
amplitude in the gauge $S_{i}=0$ is scale independent and is equal to
\[
^{(V)}h_{ij}\sim F_{i,j}\sim m.
\]
These results are valid only in linearized theory. While the result for the
tensor fluctuations remains the same, we will show in what follows that the
scalar and vector modes reach the strong coupling regime at the energy scale
which is much below the Planck scale.

\section{Vainshtein scale and continuous limit}

Let us first consider how the static interaction between two massive bodies is
modified in the Higgs model with massive graviton. In quantum field theory
this interaction is interpreted as due to the exchange by gravitons with
corresponding quantum propagators. This interpretation is very obscure from
the physical point of view because the Newtonian force is not directly related
to the propagation of gravitons. It is, however, the price to be paid in order
to preserve explicit Lorentz invariance of the theory. In our approach one
does not need to go to quantum theory to answer this question. The interaction
is entirely due to the static potentials $\phi$ and $\psi$ which are present
due to the massive body. Let us take the Newtonian gauge \cite{vm}, where
$B=E=0$ so that the metric takes the form%
\begin{equation}
ds^{2}=\left(  1+2\phi\right)  dt^{2}-\left(  1-2\psi\right)  \delta
_{ik}dx^{i}dx^{k} \label{26}%
\end{equation}
First we have to derive the equations that this metric should satisfy in
massive gravity. We consider only static solutions so all time derivatives
vanish and action (\ref{13}) simplifies to%
\begin{align}
^{(S)}\delta_{2}S  &  =\int d^{4}x\,\left\{  \psi_{,i}\psi_{,i}+\phi\left[
2\Delta\psi-m_{g}^{2}(3\psi+\Delta\pi)-T^{00}\right]  \right. \nonumber\\
&  \left.  +m_{g}^{2}\left[  3\psi^{2}+2\psi\Delta\pi+\tfrac{1}{4}\chi
_{,i}^{0}\chi_{,i}^{0}\right]  \right\}  . \label{27}%
\end{align}
We have added a term which describes the interaction with an external source
of matter for which only the $T^{00}$ component of the energy momentum tensor
does not vanish. Varying this action with respect to $\phi,\psi,\chi^{0}$ and
$\pi$ we arrive to the following equations:%
\begin{equation}
\Delta\psi=\frac{m_{g}^{2}}{2}\left(  3\psi+\Delta\pi\right)  +\frac{T^{00}%
}{2},\text{ \ \ }\Delta\left(  \psi-\phi-m_{g}^{2}\pi\right)
=0,\text{\ \ \ \ } \label{28a}%
\end{equation}%
\begin{equation}
\Delta\chi^{0}=0,\text{ \ \ \ }\Delta\left(  2\psi-\phi\right)  =0.
\label{29a}%
\end{equation}
It immediately follows from (\ref{29a}) that $\chi^{0}=0$ and $\psi=\phi/2,$
while equations (\ref{28a}) simplify to%
\begin{equation}
\Delta\left(  \phi+\psi\right)  =3m_{g}^{2}\psi+T^{00}, \label{30}%
\end{equation}
or taking into account that $\psi=\phi/2$ we obtain%
\begin{equation}
\left(  \Delta-m_{g}^{2}\right)  \phi=\frac{4}{3}\left(  \frac{T^{00}}%
{2}\right)  . \label{31}%
\end{equation}
For the central source of mass $M_{0}$ the solution of this equation is%
\begin{equation}
\phi=-\frac{4}{3}\frac{M_{0}}{r}e^{-m_{g}r}=\frac{4}{3}\phi_{N}e^{-m_{g}r},
\label{32}%
\end{equation}
where $\phi_{N}=-M_{0}/r$ is the Newtonian gravitational potential. At scales
$r\ll1/m_{g}$ the metric takes the form%
\begin{equation}
ds^{2}=\left(  1+\frac{4}{3}\left(  2\phi_{N}\right)  \right)  dt^{2}-\left(
1-\frac{4}{3}\phi_{N}\right)  \delta_{ik}dx^{i}dx^{k}. \label{33}%
\end{equation}
The bending of light is determined by the $\phi+\psi$ combination of the
metric components$.$ In General Relativity, where $\psi=\phi_{N},$ this
combination is equal to $2\phi_{N}.$ In the case of massive gravity%
\begin{equation}
\phi+\psi=\frac{4}{3}\phi_{N}+\frac{2}{3}\phi_{N}=2\phi_{N}, \label{34}%
\end{equation}
i.e. we obtain the same prediction for the bending of light. However, the
gravitational potential $\phi$ which, for instance, determines the motion of
planets has increased by factor $4/3$ compared to the Newtonian potential,
independently of the mass of the graviton. This extra contribution survives
even in the limit of zero mass. If one would redefine the gravitational
constant to get the correct Newtonian potential then obviously the bending of
light would be wrong. This is a manifestation of vDVZ discontinuity, which in
quantum field theory is interpreted as due to the propagation of the extra
scalar mode in addition to the two tensor degrees of freedom. Because this
scalar mode is coupled to the trace of the matter the result remains unchanged
for photons, but changes by the corresponding factor for non-relativistic
matter. Note that we have re-derived this result in a purely classical theory
without any reference to the tensor degrees of freedom or the
\textquotedblleft true\textquotedblright\ graviton.

The paradox with vDVZ discontinuity, which implies that the graviton must be
strictly massless was resolved when Vainshtein found a new scale $R_{V}$ in
massive gravity and suggested that for $r<R_{V}$ the scalar mode decouples and
General Relativity is restored.

We will now show how this happens in our theory, and prove that General
Relativity is smoothly restored below the Vainshtein scale. For that we will
need to consider the higher order corrections to the action (\ref{27}). First
of all we notice that because the gravitational potentials with which we are
dealing are always much smaller than unity, we can safely ignore the terms of
order $\phi^{3},\phi\psi^{2}$ etc. compared to $\phi^{2},...$ because they
cannot change the solutions of the equations drastically. We will also ignore
the terms $\phi^{2}\left(  \Delta\pi\right)  $ compared to $\phi\left(
\Delta\pi\right)  $ etc. because they are subdominant. Therefore, the only
contribution to the higher order corrections which we will take into account
will come purely from the matter scalar fields. In addition we will skip all
terms with $\chi^{0}$ since they vanish in the leading order. Hence, the only
relevant terms of the third order, which should be added to the action
(\ref{27}) are:%
\begin{align}
^{(S)}\delta_{3}S  &  =m_{g}^{2}\int d^{4}x\left[  \,\frac{1}{2}\left(
\Delta\pi\pi_{,ik}\pi_{,ik}-\pi_{,ki}\pi_{,ij}\pi_{,jk}\right)  +\frac{3}%
{16}\left(  \Delta\pi\right)  ^{3}-\frac{1}{2}\left(  \phi+2\psi\right)
\pi_{,ik}\pi_{,ik}\right. \nonumber\\
&  \left.  +2\psi\left(  \Delta\pi\right)  ^{2}+\frac{9}{16}\left(  3\psi
-\phi\right)  \left(  \Delta\pi\right)  ^{2}+O\left(  \psi^{3},\psi^{2}%
\phi,\psi^{2}\Delta\pi,\phi\psi\Delta\pi...\right)  \right]  . \label{36}%
\end{align}
These third order corrections modify the equations obtained by variation with
respect to $\psi$ and $\pi$ in the following way:%
\begin{equation}
\Delta\left(  \psi-\phi-m_{g}^{2}\pi\right)  +m_{g}^{2}\left[  \frac{3}%
{2}\left(  \phi-2\psi\right)  +\frac{1}{2}\pi_{,ik}\pi_{,ik}-\frac{59}%
{64}\left(  \Delta\pi\right)  ^{2}\right]  =0, \label{37}%
\end{equation}
and%
\begin{align}
&  \Delta\left(  2\psi-\phi\right)  +\left(  \Delta\pi\pi_{,ik}\right)
_{,ik}+\frac{1}{2}\Delta\left(  \pi_{,ik}\pi_{,ik}\right)  -\frac{3}{2}\left(
\pi_{,ij}\pi_{,jk}\right)  _{,ik}\ \nonumber\\
&  \quad\quad+\frac{9}{16}\Delta\left(  \Delta\pi\right)  ^{2}+O\left(
\phi_{,ik}\pi_{,ik},\Delta\psi\Delta\pi,...\right)  \ =0. \label{38}%
\end{align}
Equation (\ref{38}) is the main equation where non-linearities begin to play
an important role allowing us to avoid the condition $\Delta\left(  2\psi
-\phi\right)  =0,$ and thus resolve the problem of vDVZ discontinuity. In
fact, this condition means that the scalar perturbations of the curvature must
vanish, $\delta R=0,$ and this was the main obstacle leading to the troubles
with restoring General Relativity in the limit of vanishing graviton mass in
the paper \cite{boul}. \ Assuming that $\pi_{,ik}$, $\Delta\pi\ll1$ (this
assumption will be checked a posteriori), and keeping only the leading terms
in equations (\ref{37}) and (\ref{38}) we obtain
\begin{equation}
\Delta\left(  \psi-\phi-m_{g}^{2}\pi\right)  =0,\text{ \ \ \ }\Delta\left(
2\psi-\phi\right)  +\partial^{6}\pi^{2}=0, \label{39}%
\end{equation}
where by $\partial^{6}\pi^{2}$ we denoted all quadratic $\pi$ terms in
(\ref{38}). Using the first equation in (\ref{39}) to solve for $\Delta\phi,$
the second one simplifies to%
\begin{equation}
\Delta\left(  \psi+m_{g}^{2}\pi\right)  +\partial^{6}\pi^{2}=0. \label{39a}%
\end{equation}
Taking into account that $\Delta\sim\partial^{2}$ and estimating $\partial
^{6}\pi^{2}$ in spherically symmetric field $\ $as $O\left(  1\right)  \pi
^{2}/r^{6},$ this equation becomes%
\begin{equation}
\psi+m_{g}^{2}\pi+O\left(  1\right)  r^{-4}\pi^{2}\simeq0, \label{40}%
\end{equation}
The behavior of $\pi$ as a function of $r$ crucially depends on whether the
second or third term in this equation is dominating. To estimate the scale
when both terms are comparable, which is called the Vainshtein scale $R_{V},$
we set%
\[
m_{g}^{2}\pi\sim O\left(  1\right)  r^{-4}\pi^{2}\sim\psi,
\]
and from here find that%

\begin{equation}
-\left.  \psi\right\vert _{r=R_{V}}=m_{g}^{4}R_{V}^{4}.\label{42}%
\end{equation}
In the case of a gravitational field produced by the mass $M_{0}$ in the
vacuum $\psi\simeq-$ $M_{0}/r$, the Vainshtein scale is equal to
\begin{equation}
R_{V}\simeq\left(  \frac{M_{0}}{m_{g}^{4}}\right)  ^{1/5}.\label{43}%
\end{equation}
For $r\,\gg R_{V}$ the last term in (\ref{40}) is small compared to the second
one and we obtain
\begin{equation}
\pi=\frac{\psi}{m_{g}^{2}}\left[  -1+\mathcal{O}\left(  \left(  \frac{R_{V}%
}{r}\right)  ^{5}\right)  \right]  .\label{44}%
\end{equation}
In this limit the quadratic terms in the second equation in (\ref{39}) are
negligible and from the first equation in (\ref{39}) we find that%
\begin{equation}
\psi-\phi=-\psi\left[  1-\mathcal{O}\left(  \left(  \frac{R_{V}}{r}\right)
^{5}\right)  \right]  .\label{44a}%
\end{equation}
This implies that in the leading order $\psi=\phi/2$ in complete agreement
with the result which we have obtained above in linearized massive gravity. It
is easy to check that the condition $\partial^{2}\pi\ll1$ which we have used
to simplify equations (\ref{37}) and (\ref{38}) is also satisfied. In fact,
\begin{equation}
\partial^{2}\pi\sim-\frac{\psi}{r^{2}m_{g}^{2}}\sim\frac{M_{0}}{r^{5}m_{g}%
^{4}}r^{2}m_{g}^{2}\sim\left(  \frac{R_{V}}{r}\right)  ^{5}\left(  \frac
{r}{1/m_{g}}\right)  ^{2},\label{45}%
\end{equation}
and hence $\partial^{2}\pi\ll1$ for all $r\,>R_{V}$ if $R_{V}\ll1/m_{g}$.

At scales smaller than Vainshtein radius, that is for $r\,\ll R_{V}$ the third
term in (\ref{40}) is larger than the second one and hence%
\begin{align}
\pi &  \simeq O\left(  1\right)  r^{2}\sqrt{-\psi}\left[  1+O\left(  1\right)
\frac{m_{g}^{2}r^{2}}{\sqrt{-\psi}}+...\right] \nonumber\\
&  \simeq O\left(  1\right)  \frac{\psi}{m_{g}^{2}}\left(  \frac{r}{R_{V}%
}\right)  ^{5/2}\left[  1+O\left(  1\right)  \left(  \frac{r}{R_{V}}\right)
^{5/2}+\dots\right]  . \label{46}%
\end{align}
Using this expression in the first equation of (\ref{39}) we then find that in
the leading order
\begin{equation}
\psi-\phi=O\left(  1\right)  \psi\left(  \frac{r}{R_{V}}\right)  ^{5/2}+...
\label{47}%
\end{equation}
For $r\ll R_{V}$ we find that $\psi=\phi$ up to corrections of order
$\psi\left(  r/R_{V}\right)  ^{5/2}.$ Because of $\partial^{2}\pi\sim
\sqrt{-\psi}$ the condition $\partial^{2}\pi\ll1$ is always satisfied. The
dominating quadratic corrections to equation (\ref{30}) is of order $m_{g}%
^{2}\psi\sim m_{g}^{2}\left(  \partial^{2}\pi\right)  ^{2}$ so they change
only the mass term which is irrelevant within the Vainshtein scale. Taking
into account that $\psi=\phi$ for $r\ll R_{V}$ and neglecting the mass term,
equation (\ref{30}) in the leading order is reduced to
\begin{equation}
\Delta\phi=\frac{T^{00}}{2}, \label{48}%
\end{equation}
and thus General Relativity is restored within Vainshtein scales up to the
corrections%
\begin{equation}
\frac{\delta\phi}{\phi}\sim\left(  \frac{r}{R_{V}}\right)  ^{5/2}, \label{49}%
\end{equation}
which are much smaller than the corresponding corrections in DGP model
\cite{dvali2}. One could ask whether any higher order corrections would be
able to spoil the obtained results? The most dangerous of these corrections in
every next order will come as the corrections to the previous order multiplied
by $\partial^{2}\pi\ll1.$ Therefore they are completely negligible.

We now consider the implementation of our results derived classically in
quantum field theory. In the explicitly Lorentz invariant approach the change
of the interaction strength at scales exceeding the Vainshtein scale is
interpreted as due to exchange by the scalar mode $\psi$ of the massive
graviton in addition to the two tensor modes of the massless graviton. As we
will argue, this scalar mode becomes strongly coupled below Vainshtein scale
and as a result completely decouples from the gravity and matter entering the
confinement regime. This is similar to QCD, where the \textquotedblleft
soft\textquotedblright\ modes do not participate in the interactions of highly
energetic quarks below the confinement scale. Although for quantum
fluctuations one cannot neglect the time derivatives as in the static case, we
can, however, estimate the time derivatives to be of the same order of
magnitude as spatial derivatives and use the formulae derived for the static
case. Keeping in mind that the amplitude of the scalar quantum fluctuations at
the length scale $\lambda$ is about $\psi\simeq1/\lambda$ from (\ref{42}) we
obtain that at scales smaller than%
\[
\Lambda_{s}\simeq m_{g}^{-4/5},
\]
these scalar modes should be in the strong coupling regime, where nonlinear
corrections cannot be neglected. Note that the metric fluctuations which are
of order $\psi\sim m_{g}^{4/5}$ still remain small at this scale. In
distinction from the case when gravitational field is produced by an external
source the estimate $\partial^{2}\pi\sim\sqrt{-\psi}$ is not justified for
quantum fluctuations for $\lambda\ll\Lambda_{s}$. However, assuming that at
the scales which are just a bit smaller than $\Lambda_{s}$ one can still use
this estimate to find that the last term in action (\ref{13}), which is of
order $\partial^{4}\pi^{2}\sim\psi,$ becomes dominant compared to the terms of
order $\psi^{3/2}$ and $\psi^{2}.$ As a result the scalar mode $\psi$ loses
its linear propagator and decouples, entering the strong coupling regime where
nonlinear corrections will prevent its unbounded growth for every
$\lambda<\Lambda_{s}$ as $m_{g}\rightarrow0.$ As a result the terms
proportional to $m_{g}^{2}$ in the action (\ref{13}) will vanish and General
Relativity is smoothly restored in this limit. A similar thing happens with
the vector modes. Therefore in the limit $m_{g}\rightarrow0$ only the tensor
modes $\tilde{h}_{ik}$ with two degrees of freedom survive. They enter the
strong coupling regime at the Planckian scale. The energy scale $\Lambda
_{s}^{-1}$ should be taken as a cutoff scale for the scalar mode $\psi$ of
graviton in all diagrams where this scalar mode participates. Above this scale
our scalar fields $\pi$ and $\chi^{0}$ which were producing the extra degrees
of freedom for the massive graviton are also in the confined regime and the
symmetry is restored. These strongly coupled fields are completely decoupled
from gravity and the rest of the matter. In the case when the mass of the
graviton is of the order of present Hubble scale the cutoff scale is extremely
small of order $10^{-18}$ $eV.$ At higher energies the ghost, even if it would
exist, completely decouples. Therefore the question about ghosts at the
nonlinear level becomes irrelevant.

\section{How universal is the Vainshtein scale?}

The expression (\ref{43}) for the Vainshtein scale was derived first in the
case of Fierz-Pauli mass term which is unique in four dimensions, because only
in this case there are no ghosts propagating at the linear level. We have
obtained the same result in our Higgs model with the action (\ref{2}). It is
natural to ask whether it is the unique universal scale for all models with
Fierz-Pauli mass term or it depends on a particular nonlinear extension of
this term. Let us show that in our theory the Vainshtein scale, in fact,
depends on the nonlinear completion of the theory and determine all possible
extensions of the model which lead to different Vainshtein scales. With this
purpose we first consider instead of (\ref{2}) the following action for the
scalar fields%
\begin{equation}
S_{\phi}=\frac{M^{2}}{8}\int d^{4}x\,\sqrt{-g}\left[  12\left(  \frac{H}%
{4}-1\right)  ^{2}+4^{3}\beta\left(  \frac{H}{4}-1\right)  ^{3}-\tilde{H}%
_{B}^{A}\tilde{H}_{A}^{B}+O\left(  \left(  H-4\right)  ^{4}\right)  \right]
\label{49a}%
\end{equation}
where without loss of generality we have set the parameter of the symmetry
breaking to unity. The terms $O\left(  \left(  H-4\right)  ^{4}\right)  $ must
be taken in such a way as to avoid the appearance of other vacua, besides
$H=4.$ One can easily verify that there are infinitely many extensions of the
required type. This action, when rewritten in terms of $\bar{h}_{B}^{A}$
variables defined in (\ref{5}), with $v=1,$ takes the form%
\begin{equation}
S_{\phi}=\frac{m_{g}^{2}}{8}\int d^{4}x\,\sqrt{-g}\left[  \bar{h}^{2}-\bar
{h}_{B}^{A}\bar{h}_{A}^{B}+\beta\bar{h}^{3}+O\left(  \bar{h}^{4}\right)
\right]  , \label{49b}%
\end{equation}
where $m_{g}^{2}=M^{2}.$ For $\beta\gg1,$ the main contribution to the cubic
action (\ref{36}) is of order $\beta\left(  \Delta\pi\right)  ^{3}$ and the
second equation in (\ref{39}) is modified to%
\begin{equation}
\Delta\left(  2\psi-\phi\right)  +3\beta\Delta\left(  \Delta\pi\right)
^{2}=0. \label{49c}%
\end{equation}
Then using the first equation in (\ref{39}) and considering the spherically
symmetric case we find%
\begin{equation}
\psi+m_{g}^{2}\pi+O\left(  1\right)  \beta r^{-4}\pi^{2}\simeq0, \label{49d}%
\end{equation}
and correspondingly the Vainshtein scale in this case is%
\begin{equation}
R_{V}\simeq\left(  \frac{\beta M_{0}}{m_{g}^{4}}\right)  ^{1/5}. \label{49e}%
\end{equation}
Thus, we see that taking large enough $\beta$ in action (\ref{49a}) we can
obtain an arbitrarily large Vainshtein scale for given masses of the source
$M_{0}$ and the graviton $m_{g}.$

Next we would like to address the question whether one can obtain a smaller
Vainshtein scale compared to (\ref{43}). For that let us first consider the
action
\begin{align}
S_{\phi} &  =\frac{M^{2}}{8}\int d^{4}x\,\sqrt{-g}\left[  -6\left(  \frac
{H}{4}-1\right)  ^{2}\left(  \frac{H}{4}-3\right)  -\frac{1}{2}\tilde{H}%
_{B}^{A}\tilde{H}_{A}^{B}+\right.  \nonumber\\
&  \left.  +\frac{1}{2}\tilde{H}_{B}^{A}\tilde{H}_{C}^{B}\tilde{H}_{A}%
^{C}-\frac{1}{8}H\tilde{H}_{B}^{A}\tilde{H}_{A}^{B}+O\left(  \left(
H-4\right)  ^{4}\right)  ,\right]  \label{50a}%
\end{align}
where the terms $O\left(  \left(  H-4\right)  ^{4}\right)  $ are taken in such
a way as to avoid the vacuum at $H=12.$ Rewritten in terms of $\bar{h}_{B}%
^{A}$, action (\ref{50a}) becomes%
\begin{equation}
S_{\phi}=\frac{m_{g}^{2}}{8}\int d^{4}x\,\sqrt{-g}\left[  \bar{h}^{2}-\bar
{h}_{B}^{A}\bar{h}_{A}^{B}+\frac{1}{2}\left(  \bar{h}_{B}^{A}\bar{h}_{C}%
^{B}\bar{h}_{A}^{C}-\bar{h}_{B}^{A}\bar{h}_{A}^{B}\bar{h}\right)  +O\left(
\bar{h}^{4}\right)  \right]  ,\label{51}%
\end{equation}
where $m_{g}^{2}=M^{2}.$ It is clear that in the lowest order it reproduces
the Fierz-Pauli term, but in higher orders it is quite different from
(\ref{6}). The action (\ref{51}) concides with the action first derived in
\cite{gr1,gr} from the requirement of the absense of ghost in decoupling
regime up to the third order. If we consider the case of the static
gravitational field we find that in the third order the action does not
contain terms of the form $\partial^{6}\pi^{3}.$ Hence, by keeping only the
leading terms we find that the second equation in (\ref{39}) will be modified
to
\begin{equation}
\Delta\left(  2\psi-\phi\right)  +\partial^{8}\pi^{3}=0.\label{52}%
\end{equation}
Considering the spherically symmetric case and using the first equation in
(\ref{39}), which is still valid up to the leading order, we find that
equation (\ref{40}) has to be replaced by%
\begin{equation}
\psi+m_{g}^{2}\pi+O\left(  1\right)  r^{-6}\pi^{3}\simeq0.\label{53}%
\end{equation}
The Vainshtein scale will be determined by the condition that all three terms
in this equation become comparable, that is,%
\begin{equation}
\psi\sim m_{g}^{2}\pi\sim r^{-6}\pi^{3}\label{54}%
\end{equation}
and hence the expression determining this scale is%
\begin{equation}
-\left.  \psi\right\vert _{r=R_{V}}=m_{g}^{3}R_{V}^{3}.\label{55}%
\end{equation}
In particular, in the case of static field produced by mass $M_{0},$ we have%
\begin{equation}
R_{V}\simeq\left(  \frac{M_{0}}{m_{g}^{3}}\right)  ^{1/4}.\label{56}%
\end{equation}
To obtain the correction to the Newtonian potential at $r\ll R_{V}$ we note
that at these scales $\pi\sim r^{2}\psi^{1/3}$ and use of the first equation
in (\ref{39}) leads to%
\begin{equation}
\frac{\delta\phi}{\phi}\sim\left(  \frac{r}{R_{V}}\right)  ^{8/3}.\label{57}%
\end{equation}
If we set the mass of the source in (\ref{56}) to be equal to the Planck mass,
the corresponding cutoff scale in quantum theory for the decoupling of the
scalar mode is obtained : $\Lambda_{s}=m_{g}^{-3/4}.$

In principle, there are enough different combinations of $\bar{h}_{B}^{A}$
which can be added to the action (\ref{51}) to remove all the terms of the
form $\left(  \partial^{2}\pi\right)  ^{k}$ for all $k<n,$ so that the first
survived terms of this structure are $\left(  \partial^{2}\pi\right)  ^{n.}.$
Notice that such action is unique up to the order $\bar{h}^{n}.$ In this case,
the Vainshtein scale is determined by the condition%
\begin{equation}
-\left.  \psi\right\vert _{r=R_{V}}=\left(  m_{g}R_{V}\right)  ^{\frac
{2\left(  n-1\right)  }{n-2}}.\label{58}%
\end{equation}
In the case of static gravitational field due to a massive source $M_{0}$ this
yields%
\begin{equation}
R_{V}=\left(  M_{0}^{n-2}m_{g}^{2\left(  1-n\right)  }\right)  ^{\frac
{1}{3n-4}}\label{59}%
\end{equation}
and the correction to the gravitational potential for $r\ll R_{V}$ \ is of
order%
\begin{equation}
\frac{\delta\phi}{\phi}\sim\left(  \frac{r}{R_{V}}\right)  ^{\frac{3n-4}{n-1}%
}\label{60}%
\end{equation}
in agreement with \cite{dhk}. In the limit when $n\rightarrow\infty$ the
Vainshtein scale is $R_{V}=M_{0}^{1/3}m_{g}^{-2/3}.$ It coincides with the
corresponding scale in the DGP model. However, the corrections to the
gravitational potential which decay as $\left(  r/R_{V}\right)  ^{3}$ seem
different. In this limit the theory is unambiguous, but one could write it
only as an infinite series. In turn this indicates that such theory is most
probably nonlocal. Moreover, because $\partial^{2}\pi\rightarrow1$ we
completely lose control of higher order corrections and hence the results
become completely unreliable.

\section{Conclusions}

We have addressed the most fundamental question of all theories of massive
gravity - can massive gravity be a consistent theory not contradicting to
current experimental and theoretical knowledge? In this paper we have treated
gravity mostly as a classical field theory and have explicitly investigated
the issue of a smooth limit of massive gravity to General Relativity. With
this purpose we first determined the physical degrees of freedom of the
massive graviton generated via Higgs mechanism. This was done in the framework
of irreducible representations of the three dimensional rotation group, where
the five degrees of freedom of the graviton are described in terms of a tensor
mode with two degrees of freedom and vector and scalar perturbations due to
the scalar fields. The propagator for each of these five constituents of
massive gravity was derived separately. In the linear approximation the origin
of the well-known vDVZ discontinuity at the zero mass limit was traced to the
constraint equations and it was shown how the scalar and vector modes of
metric perturbations become non-dynamical in this limit.

It has been suggested long ago that the linear perturbation theory of massive
gravity fails at length scales below the Vainshtein scale and one has to
consider the full nonlinear theory to recover General Relativity below this
scale. We have determined the Vainshtein scale in Higgs gravity, with
Fierz-Pauli mass term, and found the explicit solution for the spherically
symmetric gravitational field. We have shown that the massive gravity solution
outside the Vainshtein scale smoothly goes to the General Relativity solution
in the region deep inside the Vainshtein scale. Thus the classical results and
predictions of General Relativity are recovered inside the Vainshtein scale
and at distances exceeding the Vainshtein radius, massive gravity strongly
differs from Einstein theory. This means that the scalar mode of massive
graviton decouples at Vainshtein scale and enters the strong coupling regime.
In the limit of vanishing mass, \ when Vainshtein radius becomes infinite, the
symmetry is restored and our theory is reduced to General Relativity with four
scalar fields which are confined and thus decoupled from gravity and other
matter. Based on these results we have argued that in quantum theory there is
a cutoff energy scale above which the scalar fields responsible for the scalar
and vector modes of the massive graviton are strongly coupled and confined and
hence harmless. For the realistic graviton mass this scale is extremely low.
Therefore, the question about extra scalar mode and ghost instability seems to
be irrelevant in our model.

We have found how the Vainshtein scale depends on the particular Higgs model
or, in other words, on the nonlinear extension of the Fierz-Pauli mass term.
In particular, we have shown that for given masses of the graviton and source,
the Vainshtein length scale depends on the Lagrangian of the scalar fields and
can be made arbitrary large. On the other hand, we have also constructed
Lagrangians, which produce smaller scales compared to the standard one.
However, the smallest possible scale seems to be larger than $M_{0}^{1/3}%
m_{g}^{-2/3}.$

Finally, we have calculated the corrections to General Relativity within the
Vainshtein scale which could, in principle, be interesting from experimental
point of view.

\begin{acknowledgement}
\textit{We are grateful to E.Babichev and A. Gruzinov for helpful discussions.
The work of AHC is supported in part by the National Science Foundation grant
0854779. L.A and V.M. are supported by TRR 33 \textquotedblleft The Dark
Universe\textquotedblright\ and the Cluster of Excellence EXC 153
\textquotedblleft Origin and Structure of the Universe\textquotedblright.}
\end{acknowledgement}

%\textit{\pagebreak\bigskip}

\end{document}